\documentclass[newstyle,twocolumn,proceedings]{rmaa}
\usepackage{rmaacite}

\title{Galaxies at $z=3$ around Damped Ly-$\alpha$ clouds}

\author{N. Bouch\'e\altaffilmark{1,2} \& J. D. Lowenthal \altaffilmark{3}
  \altaffiltext{1}{University of Massachusetts-Amherst, MA} 
  \altaffiltext{2}{EARA Marie-Cure fellow at the Institute of Astronomy,
  Madingley Rd, Cambridge,UK}
  \altaffiltext{3}{Smith College, Northampton, MA}
  }

\fulladdresses{
\item N. Bouch\'e: University of Massachusetts, Department of Astronomy,
LGRT 619, Amherst, MA 01003 (bouche@astro.umass.edu)
\item J. D. Lowenthal: Smith College, Department of Astronomy, Northampton, MA (james@earth.ast.smith.edu)

}
    
\shortauthor{BOUCH\'E \& LOWENTHAL} \shorttitle{Galaxies at $z=3$ around DLAs}

\ReceivedDate{1999 May 26} 
\AcceptedDate{2000 April 13} 
\SetYear{2002}
\abstract{
We are exploring the  connection between
damped Ly-alpha absorbers (DLAs) s and Lyman break galaxies (LBGs) using deep -- m$_{lim,I}$(5$\sigma$)=26 m$_{AB}$-- broad band
imaging (UBVI) of four wide fields (0.25deg$^2$ each)  obtained  at the Kitt Peak 4-m telescope with MOSAIC.
Each field contains a DLA at $z\sim3$.
	
	We want to address the nature of DLAs
at high-redshifts: (1)  Are they embedded in much larger systems of galaxies? (2) How does the 
spatial distribution of LBGs in 3D (space and redshift) correlate with the absorber? 
Contrary to most previous DLA studies, we are not looking for the absorber, and we do not 
rely on control fields because each of our fields is  $40\times 40 h^{-1}$ Mpc (co-moving).
We present preliminary results in two of our fields. In one case, we see an indication of
an overdensity of galaxies on a scale of 5 Mpc. 
We discuss the possible implications and sources of contamination of our results.

}

 %, Q2342+3417, J0124+0044)} 

\begin{document}
\maketitle
%\tableofcontents

\section{Introduction}
\label{sec:intro}
	By definition, damped Ly-$\alpha$ absorbers (DLAs) found in the spectra of quasars have neutral hydrogen
(HI) column densities greater than $2\times 10^{20}$ cm$^{-2}$. Their nature is unknown and 
has been an ongoing debate for more than a decade. Two classes of hypothesis can explain DLA properties,
(1) DLAs are large disks or proto-disks, or (2) DLAs arise from the superposition of small gas clouds
in a large halo. The former was first proposed by Wolfe et al. (1986), the latter is more recent.
Maller et al. (2000) showed that DLAs can arise from the combined effects of massive central galaxies and
a number of smaller satellite galaxies in a virialized halo. In this scenario, a DLA would lie in an over dense
region. 

	Lyman Break Galaxies (LBGs) are actively star forming objects, hence are bright $L_{UV}=10\times L^*(today)$
(e.g. Steidel 2000)  and they
 trace the underlying matter distribution as indicated by their strong clustering.

 	Based on this, the motivation of this work is to try to answer the following
question: {\it Are DLAs in over- or under-dense regions?} 
In order to test this, we selected four QSO fields that have (i) a low galactic extinction, and (ii)
$z_{DLA}\sim 3 < z_{QSO}$. In order to constrain the environment and the
nature of the damped systems at $z\sim3$, we imaged four fields with the MOSAIC camera
 at the Kitt Peak 4-m telescope in UBVI. The area covered (0.25deg$^2$/field to $I_{AB}(5\sigma)<26$) allow
us to identify several hundred LBG candidates both close to the QSO/DLA line of sight (50 kpc) and up
to 20 $h^{-1}$ Mpc (co-moving) away. Thus, unlike previous studies, we do not rely on a randomly selected
control field. 

	In this proceeding, we present preliminary results in the fields APM 08279+5255 and PC1233+4752
that contain a DLA at $z_{DLA}=2.97$ and $z_{DLA}=3.5$ respectively. 

 At $z\sim3$,  $1\arcmin$ corresponds to $\sim 1.3h^{-1}_{100}$~Mpc (co-moving)
for  $\Omega_M=0.3$, $\Omega_{\Lambda}=0.7$.

%\section{Data}
%\label{sec:data}

\section{Photometric Redshifts}
\label{sec:photoz}

 We used the photometric redshift code from Bolzonella et al.(2000) 
to estimate the redshifts of our candidates.
In order (1) to assess  the accuracy and reliability of the technique,
 and (2) to refine our set of templates, we
ran various tests on the HDF-N published data (see e.g. Fern\'andez-Soto 2001). 
We find that 
(1) near-IR data are not necessary in our redshift range ($2.8<z<3.8$), and
(2) our best set of templates consists of  4  Coleman Wu \& Weedman (1980) templates
(E, Sbc, Scd \& Irr) extended in the UV using
Bruzual \& Charlot (1993) models, and 4 Starburst99  (50Myr, 100Myr with 2 metallicities) from
Leitherer et al. (1999), 
(3) the rms of the normalized difference between photometric and spectroscopic redshifts $|z_{spec}-z_{sphot}|/(1+z_{spec}$
is 0.05 in our redshift range ($2.8<z<3.7$) 
and
(4) the number of low redshift objects contaminating  our high redshift sample is on the order of 10\%.
We plan to implement Bayesian's prior probabilities as shown in Benitez (2000) to significantly reduce the level of
contamination in the future.

\section{Results and Discussion}
\label{sec:spatial}
We selected high-redshift candidates to be U-band drop-outs with $21<I<24.5$. We then estimated
the redshift according to section~\ref{sec:photoz}.
Our results, the spatial and radial distributions of LBGs candidates around the DLA, are shown on 
Figure~1 for APM08279+5255 separated in  two redshift bins around the DLA, 
$|z_{DLA}-z|<0.2$, and $0.2<|z_{DLA}-z|<0.5$. The surface density indicates an overdensity.
 The upper limit is 0.5 to account for incompleteness 
in our redshift distribution for our selection function drops sharply at $z\sim2.8$. 
In each case, the radial surface density is plotted. Note that we find the same
signature even when we split the first bin in half.   
Around PC1233+4752, we do not find such a signature.

%\section{Discussion \& Future work}
%\label{sec:discussion}
%As opposed to Adelberger et al. (2002), 
Our results indicate that at least some DLAs may lie in overdense regions. 
We note that Adelberger et al. (2002) found  an under-density (marginally), and Gawiser et al. (2001)
were inconclusive. 
Our results are potentially limited by our lack of accurate redshifts
--- $\Delta z= 0.1$ corresponds to 75 $h^{-1}$ Mpc at $z=3$---  and contamination of
low redshift objects and red stars (estimated to be  10\% or less). This
would tend to reduce the overdensity signature. If our results were due
to a clustering of stars, the signature would disappear when we increase our lower
magnitude limit. We find the over-dense signature remains.

This Fall, we will obtain multi-object spectroscopy (with GMOS) of some of our candidates 
in order to test our photometric redshifts and to confirm our results.
We are in the process of applying the same techniques to our two other fields
(Bouch\'e et al 2003).

\begin{figure}
\begin{center}
\includegraphics[width=\columnwidth]{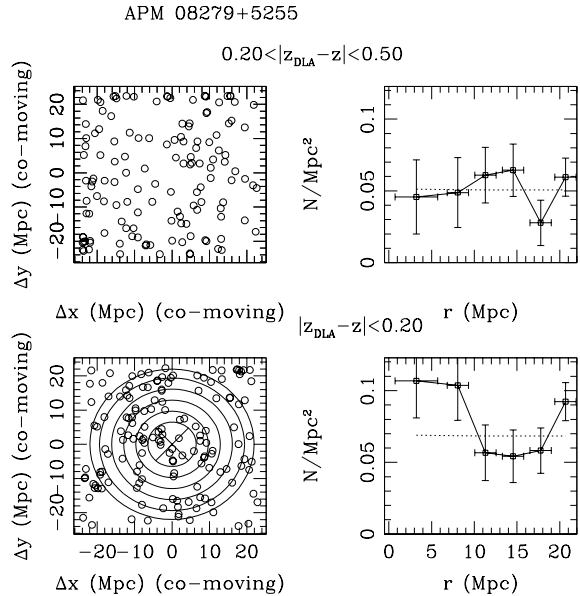}
\end{center}
\caption{Left: Spatial distribution of LBG candidates. Right: Radial surface density as a function
of radial distance from the QSO line-of-sight (shown with the cross). The lower panels are for the candidates with
$|z_{DLA}-z|<0.2$, and the upper panels are for $0.2<|z_{DLA}-z|<0.5$. The dashed line is the average
of density of the last three radial bins, representing the underlying density.
A $1\sigma$ signature of an overdensity is seen.
\label{fig1}}
\end{figure}


\begin{thebibliography}
\bibitem[Adelberger <2000>]{adelberger} Adelberger K. et al 2002 (private communication)

%\bibitem[Benitez <2000>]{benitez00} Benitez N. 2000, \apj, 536, 571 
 \bibitem[Benitez <2000>]{benitez00} Benitez N. 2000, \apj, 536, 571 

%\bibitem[Bolzonella et al.<2000>]{bolzonella} Bolzonella M., Miralles J-M. \& Pell\'o R. 2000, \aap, 363, 476
\bibitem[Bolzonella et al.<2000>]{bolzonella} Bolzonella et al. 2000, \aap, 363, 476

\bibitem[Bouch\'e et al. <2003>]{bouche03} Bouch\'e et al. 2003, in preparation

\bibitem[Bruzual \& Charlot <1993>]{bruzual93} Bruzual A.G. \& Charlot S. 1993, \apj, 405, 538 

%\bibitem[Coleman Wu \& Weedman <1980>]{coleman80} Coleman G.D., Wu, C.-C. \& Weedman, D.W. 1980, \apjs, 43, 93
\bibitem[Coleman Wu \& Weedman <1980>]{coleman80} Coleman, Wu,  \& Weedman, 1980, \apjs, 43, 93

%\bibitem[Fernandez-Soto et al.<2001>]{ferandez-soto} Fern\'andez-Soto, A., Lanzetta K.M., Chen H-W., Pascarelle S. M.
%\&  Yahata, N. 2001, \apjs, 135, 41
\bibitem[Fernandez-Soto et al.<2001>]{ferandez-soto} Fern\'andez-Soto et al. 2001, \apjs, 135, 41

\bibitem[Gawiser et al.<2001>]{gawiser} Gawiser  et al. 2001, \apj, 562, 628

%\bibitem[Leitherer et al.<1999>]{leitherer99} Leitherer C., Schaerer D., Goldader J. D., Delgado, R.M.G.,  Robert, C.,
% Kune D. F., de Mello D. F., Devost D. \&  Heckman T.M. 1999, \apjs, 123, 3
\bibitem[Leitherer et al.<1999>]{leitherer99} Leitherer et al. 1999, \apjs, 123, 3

%\bibitem[Maller et al.<2000>]{maller00} Maller, A. H.,  Prochaska, J. X., Sommerville, R. S., \&  Primack, J. R. 2000, in {\it Clustering at high redshift}, ASP vol. 200, Eds A. Mazure
\bibitem[Maller et al.<2000>]{maller00} Maller et al. 2000, in ASP vol. 200, Eds A. Mazure

%\bibitem[Pettini et al.<2000>]{pettini00} Pettini, M., Ellison, S. L., Steidel, C.C., Shapley, A. E, \&  Bowen, D.V. 2000, \apj, 532, 65  
\bibitem[Pettini et al.<2000>]{pettini00} Pettini et al. 2000, \apj, 532, 65  

%\bibitem[Rao \& Turnshek <2000>]{rao00} Rao, S. M. \& Turnshek, D. A. 2000, \apjs, 131, 1
\bibitem[Rao \& Turnshek <2000>]{rao00} Rao, S. M. \& Turnshek, D. A. 2000, \apjs, 131, 1

\bibitem[Steidel <2000>]{steidel} Steidel, C. C. 2000, SPIE, 4005, 22 

%\bibitem[Weinberg, Hernquist \& Katz <2002>]{weinberg02} Weinberg D. H., Hernquist L. \&  Katz N. 2002, \apj, 571, 15
%\bibitem[Weinberg, Hernquist \& Katz <2002>]{weinberg02} Weinberg et al . 2002, \apj, 571, 15

%\bibitem[Wolfe et al.<1986>]{wolfe86} Wolfe, A. M., Turnshek, D. A., Smith, H. E., \& Cohen, R. D. 1986,  \apjs,  61, 249
\bibitem[Wolfe et al.<1986>]{wolfe86} Wolfe et al. 1986,  \apjs,  61, 249

%\bibitem[Wolfe et al.<1995>]{wolfe95} Wolfe, A. M., Lanzetta, K. M., Foltz, C. B., \& Chaffee, F.H. 1995, \apj,  454, 698
%\bibitem[Wolfe et al.<1995>]{wolfe95} Wolfe et al. 1995, \apj,  454, 698

\end{thebibliography}
\end{document}